\documentclass[acmsmall]{acmart}

\usepackage[linesnumbered,ruled,vlined,noend]{algorithm2e}
\usepackage{tabularx}
\usepackage{multirow}
\usepackage{tablefootnote}
\usepackage{threeparttable}
\SetKwRepeat{Do}{do}{while}
\usepackage[skins]{tcolorbox}
\usepackage{xurl}
\usepackage{tikz}
\usepackage{subfigure}
\usepackage{pifont}
\usepackage{fontawesome5}
\usepackage{enumitem}
\usepackage[normalem]{ulem}

\newcommand*\circled[1]{\tikz[baseline=(char.base)]{
            \node[shape=circle,fill,inner sep=0pt] (char) {\textcolor{white}{#1}};}}

\newcommand{\myfancylabel}{\begin{tikzpicture}[every node/.style={rotate=45}]%
\node[fill,inner sep=0pt,minimum size=0.5ex] at (0ex,0.5ex) {};%
\node[fill,inner sep=0pt,minimum size=0.5ex] at (0ex,-0.5ex) {};%
\node[fill,inner sep=0pt,minimum size=0.5ex] at (0.5ex,0ex) {};%
\node[fill,inner sep=0pt,minimum size=0.5ex] at (-0.5ex,0ex) {};%
\end{tikzpicture}}

\SetCommentSty{mycommfont}
\newcommand*{\MyIndent}{\hspace*{0.3cm}}
\newcommand{\code}[1]{\texttt{#1}}

\newcommand{\rqone}{{What domains do packages in TensorFlow SC and PyTorch SC cover?}}
\newcommand{\rqtwo}{{What kinds of package clusters are formed in the two SCs?}}
\newcommand{\rqthree}{{To what extent and why do packages disengage from the two SCs?}}

\tcbset{
    enhanced,
    colframe=black!80,
    colback=black!10,
    notitle,
    top=4pt,
    left=4pt,
    right=4pt,
    bottom=4pt,
}

\AtBeginDocument{%
  }

\setcopyright{acmlicensed}
\copyrightyear{2018}
\acmYear{2018}
\acmDOI{XXXXXXX.XXXXXXX}

\acmJournal{JACM}
\acmVolume{37}
\acmNumber{4}
\acmArticle{111}
\acmMonth{8}

\begin{document}

\title{Characterizing Deep Learning Package Supply Chains in PyPI: Domains, Clusters, and Disengagement}

\author{Kai Gao}
\email{gaokai19@pku.edu.cn}
\orcid{0000-0002-0942-7890}
\affiliation{%
  \institution{School of Software \& Microelectronics, Peking University}
  \streetaddress{No. 5 Yiheyuan Road, Haidian District}
  \city{Beijing}
  \postcode{100871}
  \country{China}
}
\affiliation{%
  \institution{Key Laboratory of High Confidence Software Technologies, Ministry of Education}
  \country{China}
}

\author{Runzhi He}
\email{rzhe@pku.edu.cn}
\orcid{0000-0002-6181-6519}
\author{Bing Xie}
\email{xiebing@pku.edu.cn}
\orcid{0000-0002-2988-2575}
\author{Minghui Zhou}
\email{zhmh@pku.edu.cn}
\orcid{0000-0001-6324-3964}
\affiliation{%
  \institution{School of Computer Science, Peking University}
  \streetaddress{No. 5 Yiheyuan Road, Haidian District}
  \city{Beijing}
  \postcode{100871}
  \country{China}
}
\affiliation{%
  \institution{Key Laboratory of High Confidence Software Technologies, Ministry of Education}
  \country{China}
}

\renewcommand{\shortauthors}{K. Gao et al.}

\begin{abstract}
Deep learning (DL) frameworks have become the cornerstone of the rapidly developing DL field. 
Through installation dependencies specified in the distribution metadata, numerous packages directly or transitively depend on DL frameworks, layer after layer, forming DL package supply chains (SCs), which are critical for DL frameworks to remain competitive. 
However, vital knowledge on how to nurture and sustain DL package SCs is still lacking. Achieving this knowledge may help DL frameworks formulate effective measures to strengthen their SCs to remain competitive and shed light on dependency issues and practices in the DL SC for researchers and practitioners. 
In this paper, we explore the domains, clusters, and disengagement of packages in two representative PyPI DL package SCs to bridge this knowledge gap. 
We analyze the metadata of nearly six million PyPI package distributions and construct version-sensitive SCs for two popular DL frameworks: TensorFlow and PyTorch. 
We find that popular packages (measured by the number of monthly downloads) in the two SCs cover 34 domains belonging to eight categories. \emph{Applications}, \emph{Infrastructure}, and \emph{Sciences} categories account for over 85\% of popular packages in either SC and TensorFlow and PyTorch SC have developed specializations on \emph{Infrastructure} and \emph{Applications} packages respectively. 
We employ the Leiden community detection algorithm and detect 131 and 100 clusters in the two SCs. The clusters mainly exhibit four shapes: Arrow, Star, Tree, and Forest with increasing dependency complexity. Most clusters are Arrow or Star, while Tree and Forest clusters account for most packages (Tensorflow SC: 70.7\%, PyTorch SC: 92.9\%). 
We identify three groups of reasons why packages disengage from the SC (i.e., remove the DL framework and its dependents from their installation dependencies): dependency issues, functional improvements, and ease of installation. The most common reason in TensorFlow SC is dependency incompatibility and in PyTorch SC is to simplify functionalities and reduce installation size. 
Our study provides rich implications for DL framework vendors, researchers, and practitioners on the maintenance and dependency management practices of PyPI DL SCs. 
\end{abstract}



\received{26 June 2023}
\received[revised]{11 November 2023}
\received[accepted]{18 December 2023}

\maketitle

\section{Introduction}\label{ss:intro}
In the past decade, deep learning (DL) techniques have been widely adopted in diverse tasks such as face recognition~\cite{cvpr19-deng}, machine translation~\cite{iclr15-Bahdanau}, and code generation~\cite{aaai20-sun} with exhilarating performance. 
The burst of DL techniques is indispensable from DL frameworks, which provide a collection of APIs to allow developers to design and train DL models more easily and quickly and have become the cornerstone of the DL field. 
Many DL frameworks have been launched by various organizations, among which TensorFlow~\cite{tensorflow} and PyTorch~\cite{pytorch} are the most popular~\cite{sosurvey}. 

Research has revealed that developers have different needs when using DL frameworks to perform different tasks~\cite{tse22-gaokai}. To satisfy developers' diverse needs, substantial packages have been released by DL enthusiasts in the Python community~\cite{dlwp}. 
These packages provide a variety of specialized functionalities based on the APIs provided by DL frameworks directly or transitively. To allow package management tools like \code{pip} to automatically install DL frameworks at the installation time, DL frameworks are specified as direct or transitive \emph{installation dependencies} in these packages' distribution metadata~\cite{Coremetadata}.\footnote{According to PyPA~\cite{Glossary}, the distribution is a specific version of a package.} 
Gradually, starting from DL frameworks and through layer-by-layer installation dependencies, a large number of packages form DL package \emph{supply chains} (SCs). 

Numerous DL frameworks have emerged as DL thrives. However, remaining competitive amidst rivals is not that easy for DL frameworks. Recent years have witnessed the stop of development of several DL frameworks such as CNTK~\cite{microsoftcntk} and Chainer~\cite{chainer} for their lack of competitiveness. 
Given the wide adoption of DL techniques in diverse tasks and the prevalence of reusing third-party packages~\cite{synk}, we argue that a prosperous DL package SC is critical to a competitive DL framework. 
First, packages in the SC provide a variety of off-the-shelf specialized functionalities for developers to use, which can help DL frameworks better meet developers' diverse needs~\cite{tse22-gaokai}. 
Second, the number of dependent packages is perceived by practitioners as an indicator of the package's popularity and influences the selection of third-party packages~\cite{fse20-larios}. In this sense, a prosperous SC is helpful for DL frameworks to attract users. 
Moreover, the significance of packages in the SC is acknowledged by DL frameworks, e.g., the PyTorch website has a dedicated page presenting packages depending on PyTorch~\cite{torchEcosystem}. 
Therefore, it is important for DL frameworks to nurture and sustain their package SCs. 

Previous work has primarily studied SCs at the whole packaging ecosystem level rather than the specific field level like DL. These studies constructed SC in packaging ecosystems as a directed graph and designed metrics to measure its structure and evolution~\cite{msr16-wittern, msr17-kikas, saner17-decan, msr18-decan, emse19-decan, icse22-liuchengwei}. However, these metrics are built mainly from the graph's perspective and can not well reflect SC's nature, such as the domain of packages in the SC. 
A few studies investigated code repositories that declare dependencies on DL frameworks in the manifest files (i.e., installation dependencies)~\cite{icsme20-hanjunxiao} or in the import statements (i.e., \emph{import dependencies})~\cite{tosem21-Dilhara,icse22-tanxin}, e.g., application domains~\cite{icsme20-hanjunxiao, icse22-tanxin}, update behaviors~\cite{icsme20-hanjunxiao, tosem21-Dilhara}, and popularity factors~\cite{icse22-tanxin}. However, a code repository may or may not be released as a PyPI package and the import dependency differs from the installation dependency, leaving the knowledge on how to nurture and sustain DL package SCs still fragmented and incomplete. The primary significance of achieving such knowledge is to help \emph{DL framework vendors} formulate effective measures to strengthen their SCs to remain competitive (as argued above). Besides, such knowledge can also shed light on dependency management issues and practices in the DL SC for \emph{researchers} and \emph{DL practitioners}. 

To make a step towards filling this knowledge gap, this paper investigates two representative DL package SCs in the PyPI packaging ecosystem, which start from TensorFlow and PyTorch respectively (TensorFlow SC and PyTorch SC for short). TensorFlow and PyTorch are the two most widely adopted DL frameworks~\cite{tse22-gaokai} and PyPI is the official third-party package registry~\cite{PyPI} for Python, the mainstream programming language in the DL field~\cite{dlwp} and the main language for which TensorFlow and PyTorch provide APIs~\cite{tensorflowAPI, pytorchAPI}. 
Considering that a prosperous DL SC should help meet DL developers' diverse needs and requires the engagement of numerous packages, we formulate the following research questions: 

\textbf{RQ1 (Domains): } \emph{\rqone} The domain of a package refers to what kinds of tasks the package can solve. Investigating the domains of packages in the SC can pinpoint domains in which a DL SC should provide support to help meet developers' needs. Besides, it also helps us gain a better understanding of the SC's capability, e.g., its strengths, weaknesses, and common usage scenarios. Such understanding can help DL framework vendors strengthen their SCs to better meet developers' needs and help DL practitioners make more informed decisions when selecting DL framework (SC) for their tasks. Prior work has studied application domains of code repositories that directly depend on DL frameworks~\cite{icsme20-hanjunxiao} and domains of packages in the DL repository SC based on import dependencies~\cite{icse22-tanxin}. However, they cover only a small part of the DL package SC (more details in Section~\ref{s: sc construction}) and a more complete understanding of domain distributions in the DL package SC is still lacking. 

\textbf{RQ2 (Clusters):} \emph{\rqtwo} DL package SC is a complex graph involving numerous interdependent packages, which makes its structure mysterious. Decomposing DL package SC into package clusters where packages are densely connected in the same cluster but sparsely connected between clusters, is a good means to understand how packages engage in the SC and identify critical packages that attract other packages to the SC, thus helping DL frameworks take effective actions to attract packages. For example, packages on which the other packages depend in each cluster significantly impact the SC's proper functioning and should be prioritized for security measures. Besides, packages in the same cluster may share similar functionality, thus package clusters can serve as a valuable source for practitioners to search for suitable packages in the SC. 

\textbf{RQ3 (Disengagement):} \emph{\rqthree} It is common for software to remove some dependencies during its evolution~\cite{fse21-hehao}. If a package completely removes dependencies on the DL framework and its dependents since a certain version, we refer to that package disengaging from the DL package SC. The disengagement of packages may affect the SC's sustainability and stability and lead to user churn, thus reducing the DL framework's competitiveness. Moreover, the disengagement of packages could also signal potential dependency management problems faced by package developers in the SC. Therefore, understanding why packages disengage from the SCs helps DL frameworks take countermeasures to keep packages engaged. However, not much is known about the prevalence and rationale for package disengagement in the DL package SC. 

To answer the questions, we first build an ecosystem-wise and version-sensitive installation dependency database by analyzing the metadata of nearly six million PyPI package distributions. Based on the database, we construct TensorFlow and PyTorch SC by iteratively retrieving packages that declare direct or transitive installation dependency on TensorFlow or PyTorch. We conduct a thematic analysis on the description of popular packages (measured by the number of monthly downloads) in the two SCs and identify 34 package domains spanning eight categories. Over 85\% of popular packages in either SC fall into \emph{Applications}, \emph{Infrastructure}, and \emph{Sciences} categories. \emph{Infrastructure} and \emph{Applications} categories account for the most popular packages in TensorFlow SC and PyTorch SC respectively, uncovering the two SCs' different specializations. We employ the Leiden community detection algorithm and detect 131 and 100 package clusters in the two SCs. The clusters mainly exhibit four shapes: Arrow, Star, Tree, and Forest, with increasing dependency complexity. Despite most clusters being Arrow or Star, Tree and Forest clusters account for the majority (70.7\% and 92.9\% respectively) of packages, indicating that a small number of packages attract most of the other packages to the SC. We find the number of packages that have disengaged from either SC shows an increasing trend. We identify seven disengagement reasons related to three aspects: dependency issues, functional improvements, and ease of installation. The most common disengagement reasons differ between the two SCs, suggesting different dependency management problems in them. 

In a nutshell, this paper makes the following contributions: 
\begin{itemize}
    \item A comprehensive understanding of the domains, clusters, and disengagement of packages in two representative DL package SCs in the PyPI packaging ecosystem.     
    \item An automated approach to construct version-sensitive PyPI package SCs and a public dataset and scripts\footnote{\url{https://github.com/gaokai320/PyPI-DLSC}} to facilitate future research. 
    \item Rich suggestions for DL framework vendors, researchers, and DL practitioners on the maintenance and dependency management practices of PyPI DL package SCs. 
\end{itemize}

\section{Background and Related Work}\label{s: background and related work}

\subsection{Background}\label{ss: background}

\emph{\textbf{Definition of Software SC.}} 
In traditional production, SC refers to the network of companies involved in the process of converting raw materials to final products delivered to customers, where supply relationships between companies are established by the flow of products~\cite{scm}. Similar supply relationships exist between software~\cite{floss2019amreen}. Specifically, if a piece of software uses the functionality provided by another piece of software (e.g., package APIs), a supply relationship (i.e., dependency) is assumed to be established between them. As software development becomes increasingly dependent on external software, supply relationships among software become more and more complex, gradually forming a network of software connected by dependencies, i.e., software SC.

\noindent \emph{\textbf{Code Repository vs. Package Distribution}.} 
Prior work~\cite{icse22-tanxin} investigated code repositories that depend on DL frameworks and analyzed the domains of code repositories that release packages to PyPI by matching the repository's URL with the code repository URL declared in the PyPI package's distribution metadata. However, substantial (about 1/3) PyPI packages do not declare their code repository information in the distribution metadata.\footnote{The script that calculates the ratio is provided in the replication package.} 
Therefore, identifying PyPI packages from code repositories will omit considerable PyPI packages. 
In this paper, we choose \textbf{package distributions} published in PyPI since we can easily obtain a complete list of PyPI packages and dependency relationships between them, which allows us to construct as complete package SCs in PyPI as possible. 

\noindent \emph{\textbf{Installation Dependency vs. Import Dependency.}} 
Existing literature mainly investigates two kinds of dependencies: installation dependency and import dependency. 
Installation dependency is specified in the software's manifest files such as \code{setup.py}, \code{requirements.txt}, and package distribution metadata~\cite{icse20-wangying}. Import dependency is specified in the code files' import statements instead~\cite{icse22-tanxin}, e.g., \code{import torch}. 

In this paper, we choose to construct DL package SCs by analyzing \textbf{installation dependencies} between package distributions for four reasons. 
First, installation dependency is the most widely used in prior work on analyzing package SCs (e.g.,~\cite{msr16-wittern, msr17-kikas, emse19-decan, 2023ASE-Xu}). 
Second, Python has provided an official package, \code{packaging}~\cite{Packaging} which can accurately parse installation dependencies specified in the package distribution metadata. 
Third, it is challenging to accurately identify import dependencies (i.e., map the package's import name in the import statement to its package name) since a Python package's import name is not necessarily the same as its package name and different packages may share the same import name~\cite{saner23-guhaiqiao}. 
Fourth, installation dependency reserves dependency version information, which enables us to analyze packages' disengagement. 
However, installation dependency suffers from the bloated dependency issue~\cite{tse22-caoyulu}, that is, some installation dependencies are not used by the software. To evaluate the impact of this issue, we manually check some packages, as detailed in Section~\ref{s: sc construction}. 

\noindent \emph{\textbf{Definition of PyPI DL Package SC.}}
We conceptualize \emph{PyPI package SC} as a directed graph $\mathcal{G}= (\mathcal{P}, \mathcal{E})$ where vertices in $\mathcal{P}$ are PyPI packages, edges in $\mathcal{E}$ represent installation dependencies between packages. 
Formally, installation dependency can be defined as an ordered pair $(P_{u}[u_i], P_{d}[d_j])$, indicating that to install version $d_j$ of the package $P_d$, version $u_i$ of the package $P_u$ should be installed first. We refer to $P_u$ as $P_d$'s dependency and $P_d$ as $P_u$'s dependent. 
We refer \emph{PyPI DL package SC} as a sub-graph of PyPI package SC which starts from a DL framework and includes all packages directly or transitively depending on the framework, as detailed in Section~\ref{s: sc construction}.

\subsection{Related Work}
Our work lies in the intersection between software SC and DL and hence has two groups of related work: research on SCs in packaging ecosystems and research on software engineering (SE) for DL. 

\subsubsection{Research on SCs in Packaging Ecosystems} 
Due to the prevalence of reusing third-party packages in software development~\cite{fse21-hehao}, package SCs in various packaging ecosystems have been a popular topic in the SE community. 
Wittern \emph{et al.}~\cite{msr16-wittern} investigated the evolution of NPM and found that the number of packages and dependency relationships among packages increased rapidly. 
Decan \emph{et al.}~\cite{ecsaw16-alexandredecan, saner17-decan, emse19-decan} and Kikas \emph{et al.}~\cite{msr17-kikas} analyzed the distribution of direct and transitive dependencies in package SCs in multiple packaging ecosystems. They also found growing packages and dependency relationships among packages in these SCs, and that transitive dependencies lead to the fragility of these SCs. 

Substantial work is conducted to identify and mitigate risks in package SCs. 
Valiev \emph{et al.}~\cite{fse18-maratvaliev} found that PyPI packages' relative position in the SC has a significant impact on their sustainability. 
Vu \emph{et al.}~\cite{fse21-vu} investigated the differences between the code in PyPI packages and their corresponding code repositories to identify potentially malicious code injections. Liu \emph{et al.}~\cite{icse22-liuchengwei} investigated the vulnerability propagation and propagation evolution in NPM. Zahan \emph{et al.}~\cite{icse22-zahan} proposed six security weakness indicators to identify potential SC attacks in NPM. 
Wang \emph{et al.} investigated dependency conflicts introduced by transitive dependencies in PyPI~\cite{icse20-wangying} and NuGet~\cite{icse22-lizhenming}. 

Despite these efforts, knowledge of how to nurture and sustain DL package SCs is still limited. 

\subsubsection{Research on SE for DL}\label{sss: se4dl}
A large body of work has been devoted to SE for DL~\cite{tse22-gaokai} with the data from GitHub, Stack Overflow (SO), interviews, etc. 
Specifically, DL bugs are extensively investigated. Zhang \emph{et al.}~\cite{issta18-zhangyuhao} explored the symptoms, root causes, and fix challenges of bugs in TensorFlow applications. Islam \emph{et al.}~\cite{fse19-islam, icse20-Islam} and Humbatova \emph{et al.}~\cite{icse20-Humbatova} studied bugs in applications of several DL frameworks. Bugs associated with DL jobs~\cite{icse20-zhangru}, DL frameworks~\cite{dsaa20-lijia, jss21-lijia, ist22-yangyilin}, DL dependency stack~\cite{2023FSE-Huang}, training~\cite{icse21-zhangxiaoyu}, and deployment~\cite{icse21-chenzhenpeng} are also explored. 
Some research identified challenges for developing~\cite{issre19-zhangtianyi, emse20-hanjunxiao, tse22-gaokai} and deploying~\cite{fse20-chenzhenpeng} DL applications. 
This line of work provides insights for the SE community to tackle DL problems but none touches the problem from SC's perspective. 

Little work investigated the characteristics of code repositories depending on DL frameworks. 
Han \emph{et al.}~\cite{icsme20-hanjunxiao} studied the project purposes, application domains, update behaviors, and dependency version distribution of code repositories that declare DL frameworks as dependencies in the manifest files. 
Dilhara \emph{et al.}~\cite{tosem21-Dilhara} investigated the usage and update of DL frameworks in code repositories that import them in the code. 
The two studies suffer from two major limitations. First, they only focused on code repositories that directly depend on the DL frameworks and ignored the network structure of SC. Second, the code repositories they collected may or may not be released as PyPI packages. Therefore, the two studies cover only a small part of the DL package SC. 
Tan \emph{et al.}~\cite{icse22-tanxin} constructed TensorFlow and PyTorch repository SCs by repeatedly retrieving code repositories that directly or transitively import TensorFlow or PyTorch. They located code repositories released as PyPI packages in the SC and investigated the two SCs' structure, application domains, and popularity factors. 
However, their SCs differ a lot from package SCs for two reasons. 
First, their SCs miss substantial PyPI packages as discussed in Section~\ref{ss: background}. About 70\% of packages in our SCs are not covered by theirs (more details in Section~\ref{s: sc construction}). 
Second, import dependencies between the packages' code repositories may over-represent installation dependencies between package distributions. Specifically, not every file (e.g., test files and example files) in the repository will be packaged into the distribution, so import dependencies declared in these files are not installation dependencies of the distribution. A manual inspection of 100 packages in the SCs of~\cite{icse22-tanxin} indicates that the import dependencies declared in the code repository of 27 packages are not installation dependencies of their distributions. The two differences lead to the differences in the findings of domain distribution between their study and ours, as discussed in Section~\ref{ss: rq1 results}. 

To conclude, earlier studies cover only a small part of the PyPI DL package SC and their results can not sufficiently reveal how to nurture and sustain PyPI DL package SCs.\footnote{In the following sections, if not specified, ``SC'' refers to PyPI package SC, and ``dependency'' refers to installation dependency.} 

\section{Supply Chain Construction}\label{s: sc construction}
The Python Packaging Authority (PyPA), a working group responsible for Python packaging, hosts a public dataset on BigQuery containing the metadata of all distributions released on PyPI~\cite{bigquery}, which enables us to construct as complete PyPI DL package SCs as possible. We download the data dump on Nov 4, 2021, which consists of 354,636 packages and 5,743,721 distributions. 

We first build a dependency database $\mathcal{D}$ to enable the construction of PyPI DL SCs. The database is ecosystem-wise since it covers all PyPI packages. Compared with the approaches proposed in prior work~\cite{ecsaw16-alexandredecan, fse18-maratvaliev} on analyzing PyPI package SCs, it also takes the dependency's version information into account to obtain dependency relationships between PyPI package releases (i.e., version-sensitive), which enables us to construct finer-grained DL package SCs and investigate the package's disengagement. Specifically, the PyPI package distribution specifies its installation dependencies in the \code{requires\_dist} field in its metadata. 
Each item in the field contains a package name $P_u$ and an optional version constraint $\mathcal{C}$. 
A version constraint specifies a PEP508 version range~\cite{PEP508} for a dependency. 
We implement a version constraint parser on top of the \code{packaging}~\cite{Packaging} package, a tool released by PyPA to parse dependency specifications, to locate satisfying dependency versions. 
Specifically, for each version $u_i$ of $P_u$ released in PyPI (by querying the PyPI distribution metadata dump), we check if it satisfies $\mathcal{C}$. If satisfied, we insert a record $(P_{u}[u_i], P_{d}[d_j])$ to $\mathcal{D}$. 
Finally, $\mathcal{D}$ contains 281,443,215 records. 

\begin{figure}
    \centering
    \includegraphics[width=0.5\linewidth]{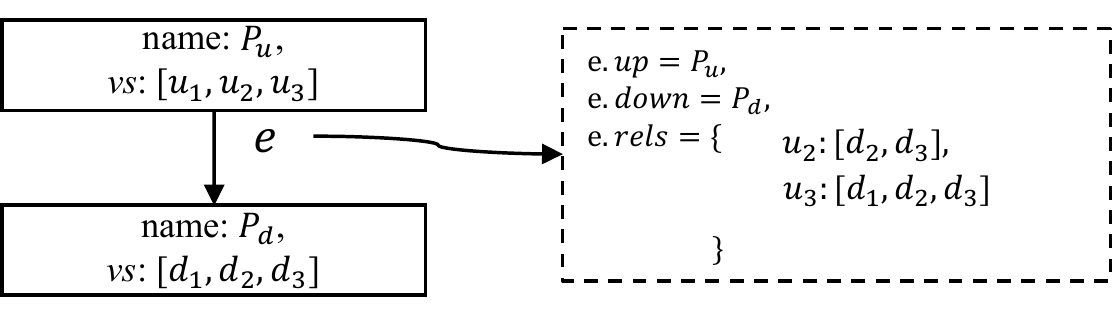}
    \caption{An example of SC.}
    \label{fig:sc example}
\end{figure}

Based on $\mathcal{D}$, we construct PyPI DL SCs as follows. 
To be version-sensitive, we add attributes to packages and dependency relationships in the SC. 
For each package, we maintain an attribute $vs$ which is a list storing all the package's \underline{v}ersion\underline{s} appearing in the SC. For each dependency relationship, we maintain an attribute $rels$ which uses a map to store the version dependency \underline{rel}ation\underline{s} between package $P_u$ and its dependent $P_d$. 
Each key in $rels$ represents a version of $P_u$ and the value represents all versions of $P_d$ that depend on the version of $P_u$ corresponding to the key. 
Take Figure~\ref{fig:sc example} as an example, suppose there is a package $P_u$ and a package $P_d$ depending on $P_u$ in the SC. 
$u_1, u_2, u_3$ versions of $P_u$ and $d_1, d_2, d_3$ versions of $P_d$ appear in the SC as indicated by the $vs$ attribute. 
The $rels$ attribute of dependency $e$ between $P_u$ and $P_d$ shows that $P_d$ only depends on the $u_2$ and $u_3$ versions of $P_u$. Specifically, the $d_2$ and $d_3$ versions of $P_d$ depend on the $u_2$ version of $P_u$ and all versions of $P_d$ depend on the $u_3$ version of $P_u$. 

\begin{figure}[ht]
  \centering
  \begin{minipage}{.75\linewidth}
    \begin{algorithm}[H]
    \caption{Constructing PyPI DL SC}
    \label{alg:sc construction}
    \DontPrintSemicolon
    \SetKwInput{Input}{Input}
    \SetKwInput{Output}{Output}
    \Input{A list of package names: $\mathcal{N}$}
    \Output{A supply chain: $\mathcal{G}$}
        $\mathcal{P} \leftarrow Vertices(), \mathcal{E} \leftarrow Edges(), unvisited \leftarrow Vertices()$\\
        \tcp{Input packages as initial unvisited packages}
        \ForEach{$n \in \mathcal{N}$}{
            $p \leftarrow Vertex(name=n, vs=list())$\\
            $p.update\_versions(GetAllPyPIVersions(n))$\\
            $\mathcal{P}.merge(p)$\\
            $unvisited.merge(p)$
        }
        \While{$unvisited \neq \emptyset$}{
            $dependents \leftarrow Vertices()$\\
            \ForEach{$u \in unvisited$}{
                \ForEach{$version \in u.vs$}{
                    $dps \leftarrow GetDependents(u.name, version)$\\
                    $dependents.merge(dps)$\\
                    \ForEach{$d \in dps$}{
                    \tcp{Merge new dependencies to $\mathcal{E}$}
                        $e \leftarrow Edge(up=u.name,down=d.name, rels=\{version: d.vs\})$\\
                        $\mathcal{E}.merge(e)$
                    }
                }
            }
            \tcp{Obtain new $unvisited$ packages and versions}
            $unvisited \leftarrow dependents - \mathcal{P}$\\
            \tcp{Merge newly found packages and versions to $\mathcal{P}$}
            $\mathcal{P} \leftarrow \mathcal{P}.merge(dependents)$
        }
        $\mathcal{G} \leftarrow DiGraph(\mathcal{P}, \mathcal{E})$\\
        \Return $\mathcal{G}$
    \end{algorithm}
\end{minipage}
\end{figure}

Based on the definition in Section~\ref{ss: background}, we iteratively construct version-sensitive PyPI DL SC following the procedure described in Algorithm~\ref{alg:sc construction}. 
Considering that a DL framework may release multiple packages, e.g., TensorFlow framework releases three packages: \code{tensorflow}, \code{tensorflow-cpu}, and \code{tensorflow-gpu}, the algorithm takes a list of package names $\mathcal{N}$ as input. 
$unvisited$ is a queue of unexplored packages. Our algorithm retrieves direct dependents of packages in $unvisited$ and updates $unvisited$ with unexplored direct dependents in each iteration. 
$unvisited$ is initialized as all input packages with all their released versions in PyPI (lines 2$\sim$6). For each version of each package in $unvisited$, we query all versions of packages that directly depend on it from $\mathcal{D}$ (lines 9$\sim$12) and merge dependencies between them to $\mathcal{E}$ (lines 13$\sim$15). The merge takes two forms: adding new edges, or updating $rels$ of existing edges (adding new key-value pairs or updating existing values). 
$dependents$ denotes all versions of packages that directly depend on packages in $unvisited$. 
We obtain new $unvisited$ packages by comparing $dependents$ with $\mathcal{P}$ (line 16) and merge newly found packages ($dependents$) to $\mathcal{P}$ (line 17). 
The merge operation either adds new packages or adds new versions to $vs$ of existing packages. 
The above process is repeated until $unvisited$ is empty (line 7). 

We construct TensorFlow and PyTorch SC with the input of \code{[tensorflow, tensorflow-cpu, tensorflow-gpu]} and \code{[torch]} respectively. 
Finally, TensorFlow SC contains 2,567 packages, 19,116 versions, and 3,232 edges; PyTorch SC contains 3,278 packages, 26,563 versions, and 5,440 edges. We also construct TensorFlow SC and PyTorch SC by Nov 2021 to make a comparison with~\cite{icse22-tanxin}. There are 1,087 unique packages in the two SCs and 772 (71.0\%) are not covered by~\cite{icse22-tanxin}. We investigate the monthly downloads and the number of dependents for the uncovered packages and find that: 1) 185 packages have above-average monthly downloads, accounting for 79.4\% of (233) packages with above-average monthly downloads; 2) 85 packages have dependents, such as \code{torchvision} with 130 dependents, \code{torchtext} with 21 dependents, \code{pytorch-transformers} with 13 dependents, and \code{tensorflow-probability} with 12 dependents. Therefore, our SCs cover more popular and critical packages than~\cite{icse22-tanxin} and provide a more complete picture of DL package SCs in PyPI.

As discussed in Section~\ref{ss: background}, the package may not use its installation dependencies (i.e., bloated dependency), affecting the accuracy of the constructed DL package SC. To evaluate the prevalence of this issue, we randomly select 334 and 344 packages from TensorFlow SC and PyTorch SC respectively (95\% confidence level, 5\% confidence interval).\footnote{\url{https://www.surveysystem.com/sscalce.htm}} For each sampled package, we download one distribution of its latest version appearing in the SC and check if this distribution imports its installation dependencies in the SC. Among the sampled packages, 7 (out of 334) and 6 (out of 344) packages have been removed from PyPI at the time of inspection. Among the remaining 327 and 338 packages, 305 (93.3\%) and 334 (98.8\%) packages import their installation dependencies, suggesting that the bloated dependency issue is rare in the two SCs we construct. Besides, it seems that the bloated dependency issue is more common in TensorFlow SC than in PyTorch SC, which coincides with the finding in Section~\ref{ss: rq3 results}. 

\section{RQ1: \rqone}

\subsection{Method}\label{ss:rq1 method} 
While manual checking is a feasible approach to identifying the package's domain, it is almost unaffordable to go through all packages due to the large number. 
We, therefore, choose to focus on popular packages in the two SCs measured by the number of monthly downloads, a widely used popularity metric~\cite{promise18-dey, fse21-vu, ase21-vu, 2023ASE-Xu}, for three reasons. 
First, intuitively, a higher number of monthly downloads indicates that the package or the package's dependents are more used by developers, which can better reflect the SC's core capabilities. 
Second, prior work~\cite{icse22-tanxin} also selected popular packages to analyze domain distribution. Third, many packages are rarely used, e.g., toy or dummy packages~\cite{EuroSPW20-Vu, icse22-tanxin}, which can not well reflect the SC's core capabilities. 
Considering newly created packages, we retrieve the number of downloads of all PyPI packages from the public PyPI download statistics dataset one month after we obtain the distribution metadata dump, i.e., between Nov 4, 2021, and Dec 4, 2021. 
Then, we select packages in the two SCs with over 3,825 monthly downloads, the average monthly downloads of PyPI packages. We use the average monthly downloads as the selection metric to balance the representativeness of selected packages and the time costs of conducting the manual inspection. In this way, we obtain 219 and 259 packages in TensorFlow and PyTorch SC respectively, which are feasible for manual labeling and are higher than the number of packages investigated in prior work~\cite{icse22-tanxin}. Perhaps not surprisingly, the popular packages in the two SCs are not disjoint with 41 overlapped packages, leaving 437 unique packages. 

We follow the general thematic analysis~\cite{esem11-cruzes} procedure to determine package domains. First, two inspectors (the first and second author) inspect each package's summary, description, and repository README to collect relevant texts about their domains. Then, they independently read and re-read all materials, generate initial codes describing each package's domain, and merge similar codes into themes. If a package can not be labeled due to the lack of information on its domain, it is marked as \emph{Unclear}. Next, they compare the lists of codes and themes until an agreement is reached and develop a coding guide with definitions and examples for each identified code. In the process, we refer to the topics listed in the call for paper of NeurIPS,\footnote{\url{https://neurips.cc/Conferences/2023/CallForPapers}} one of the top-tier conferences in the DL field to ensure the rationality of generated domains and categories. 
After that, the inspectors independently use the coding guide to label all materials, and the inter-rater reliability measured by Cohen's Kappa~\cite{cohenkappa} is 0.953. The disagreements are resolved by a discussion. 
Finally, 201 packages in TensorFlow SC and 247 packages in PyTorch SC are properly labeled. 28 packages are marked as \emph{Unclear} and are excluded in the following discussion. The entire process takes about 200 man-hours and related materials are provided in the \href{https://github.com/gaokai320/PyPI-DLSC}{replication package}. 

\begin{figure}
    \centering
    \includegraphics[width=\linewidth]{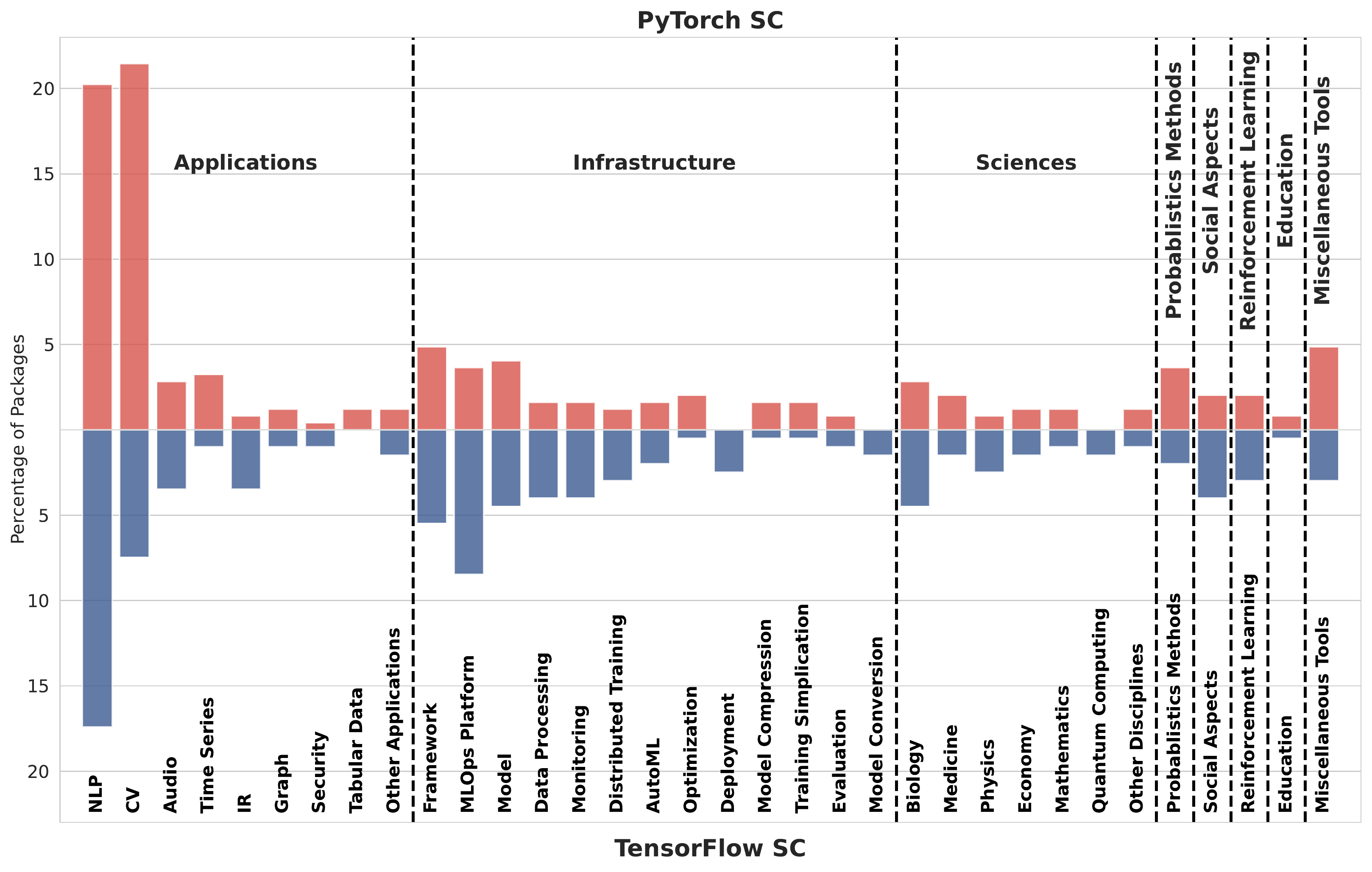}
    \caption{Distribution of the 34 package domains and eight categories. The upper part (red bars) corresponds to PyTorch SC and the lower part (blue bars) corresponds to TensorFlow SC.}
    \label{fig:package domains}
\end{figure}

\subsection{Results}\label{ss: rq1 results}
Figure~\ref{fig:package domains} presents the domain distribution covered by the popular packages in either SC. In total, we identify 34 package domains under eight categories. 

\emph{Applications} category includes nine domains that have long been studied in computer science, such as natural language processing (NLP) and computer vision (CV). Packages in this category provide functionalities that facilitate the application of DL techniques on specific types of data (e.g., images, text, and audio), such as data preprocessing and model architectures. 
This category accounts for the highest (52.6\%) number of popular packages in PyTorch SC and the second-highest (36.3\%) in TensorFlow SC, among which \textit{NLP} and \textit{CV} are predominant. 
We also notice differences between the two SCs. 
Generally, the proportion of \emph{Applications} packages in PyTorch SC is significantly higher than that in TensorFlow SC, as confirmed by the one-sided proportion z-test ($p$ < 0.001). It is known that PyTorch offers better programmabilities than TensorFlow, and our finding further suggests that PyTorch SC has developed competitive advantages in \emph{Applications} domains with many popular packages. 
Specifically, PyTorch SC provides a higher percentage of packages on \emph{NLP}, \emph{CV}, \emph{Time Series}, and \emph{Tabular Prediction}, while TensorFlow SC provides a higher percentage of \emph{IR} (Information Retrieval) packages. Besides, the number of \emph{NLP} packages is slightly less than the number of \emph{CV} packages in PyTorch SC, but in TensorFlow SC, the former is twice as many as the latter. 
Such differences may be attributed to the availability of related tools. 
For example, \code{torchvision} is the most popular package in PyTorch SC. It provides \emph{popular datasets, model architectures, and common image transformations for computer vision}~\cite{torchvision} and is dependent by 943 packages in PyTorch SC. That is, \code{torchvision} lays the foundation for the prosperity of \emph{CV} packages in PyTorch SC. But similar \emph{CV} packages are missing in TensorFlow SC. 

\emph{Infrastructure} is the second-largest (24.7\%) category in PyTorch SC and the largest (37.8\%) category in TensorFlow SC. Packages in this category provide functionalities applicable to multiple domains in other categories. It includes 13 domains. The distribution of domains in this category differs greatly between the two SCs. 
First, TensorFlow SC provides a significantly higher percentage of popular \emph{Infrastructure} packages than PyTorch SC ($p$ < 0.01), suggesting the competitive advantage of TensorFlow SC in \emph{Infrastructure} domains. 
Second, a notable difference is that TensorFlow SC has a much higher percentage of popular \emph{MLOps Platform} packages. These packages are released by various production platform vendors such as Azure Machine Learning\footnote{\url{https://azure.microsoft.com/en-us/products/machine-learning}} and TensorFlow Extended\footnote{\url{https://www.tensorflow.org/tfx}} and provide functionalities that facilitate the development, deployment, and management of DL models in these platforms. It is recognized that TensorFlow offers more features for productionization, and our finding further confirms the prosperity of productionization tools in TensorFlow SC. 
Third, the proportion of \emph{Data Preprocessing}, \emph{Monitoring}, \emph{Distributed Training}, \emph{Deployment}, and \emph{Model Conversion} packages in TensorFlow SC is higher than that in PyTorch SC but PyTorch SC provides a higher percentage of \emph{Optimization}, \emph{Model Compression}, and \emph{Training Simplification} packages. Such differences not only indicate the abundance of tool support in the SC but also reflect the functional deficiencies of the framework. For example, it is acknowledged that TensorFlow SC provides a collection of powerful deployment tools, e.g., TensorFlow Serving and TensorFlow Lite~\cite{fse20-chenzhenpeng}. However, TensorFlow does not offer native support for ONNX~\cite{onnx}, an open standard format of DL models that powers interoperability among frameworks, while PyTorch does. As a result, three popular \emph{Model Conversion} packages in TensorFlow SC emerge to enable the conversion between TensorFlow and ONNX models. Another example is the \emph{Training Simplification} packages in PyTorch SC. Despite the convenience of building complex models with PyTorch, developers have to write a lot of boilerplate code to train DL models. To address these issues, many \emph{Training Simplification} packages like \code{pytorch-ignite} are proposed to simplify intricate training configurations. 
Overall, these differences suggest that TensorFlow SC provides stronger capabilities in productionizing DL models. 

Amazingly, \emph{Sciences} category has the third most packages (13.4\% in TensorFlow SC and 9.3\% in PyTorch SC). Packages in this category facilitate the application of DL techniques in many scientific disciplines, especially \emph{Biology}, \emph{Medicine}, and \emph{Physics}. 
Next, \emph{Probabilistic Methods}, \emph{Reinforcement Learning}, and \emph{Social Aspects} take their turn. 
\emph{Probabilistic Method} packages provide various Bayesian inference methods and Bayesian optimization models, especially the Gaussian Process. PyTorch SC provides a higher percentage (3.6\%) of related packages than TensorFlow SC (2.0\%). 
\emph{Reinforcement Learning} (RL) packages provide various RL environments and models. 3.0\% of the popular packages in TensorFlow SC and 2.0\% in PyTorch SC fall into this domain. 
\emph{Social Aspects} is an emerging hot research topic aiming to reduce the adverse effects such as discrimination and unreliable decisions brought by DL. It involves many aspects such as robustness, explainability, fairness, and privacy. It seems that TensorFlow SC provides more (4.0\%) widely used packages on \emph{Social Aspects} than PyTorch SC (2.0\%), possibly due to that social aspects are of great concern when productionizing DL models and TensorFlow SC is more used to deploy DL model into production environments. 
After that, 3.0\% and 4.9\% of the popular packages in TensorFlow SC and PyTorch SC are \emph{Miscellaneous Tools}, which provide commonly used utilities and wrappers. 
Finally, we notice three popular packages in the two SCs are supporting materials for DL books and courses that teach developers DL techniques, e.g., \code{nlpia}, a package containing code examples for the \emph{Natural Language Processing in Action} book. 

We also make a comparison with the findings reported by~\cite{icse22-tanxin}, which constructed DL SCs based on import dependencies between code repositories and analyzed application domains of repositories that have released PyPI packages. For comparison, we operate as follows: 1) keep packages released before Nov 2019, the time of their data collection; 2) map \emph{Applications}, \emph{Sciences}, \emph{Probablistic Methods}, and \emph{RL} packages to domain-related (DR) packages and the rest to non-domain related (NDR) packages, based on their coding guide of package domains. 
Our study confirmed some findings by~\cite{icse22-tanxin}, e.g., TensorFlow SC provides more \emph{RL} packages; \emph{CV} and \emph{NLP} are the two most popular DR package types. 
However, there are also noteworthy differences. 
First, different from their finding that the numbers of DR and NDR packages are almost the same in either SC, we find that the number of DR packages is about 1.6 times of NDR packages in PyTorch SC, suggesting different specializations of the two SCs. 
Second, they found that there are more \emph{CV} packages than \emph{NLP} packages in either SC. In contrast, we find that the number of \emph{NLP} packages is comparable and about two times of \emph{CV} packages in PyTorch SC and TensorFlow SC, indicating a relatively greater focus on NLP in either SC. 
Third, instead of more experiment result analysis packages provided by TensorFlow SC, we find that PyTorch SC provides a higher proportion of \emph{Monitoring} packages, possibly because experiment result analysis is very important whereas PyTorch does not provide native support as TensorFlow does (e.g., the \emph{tensorboard} visualization tool). The differences can be ascribed to the differences between repository SCs and package SCs discussed in Section~\ref{sss: se4dl}. 

Among the 41 overlapped packages, 18 are about \textit{Applications} and 12 are about \textit{Infrastructure}. 
These overlapped packages depend on packages from both SCs, either providing functionalities to support both DL frameworks or using packages from one SC to complement the other. 
For example, \code{tempeh} is a framework to track the model's memory usage and run time. It depends on both TensorFlow and PyTorch to be able to support models written with TensorFlow and PyTorch. 
As an example for the latter case, \code{memcnn} is a framework built upon PyTorch and uses \code{tensorboard} from TensorFlow SC for logging since PyTorch SC did not provide related tools at that time. 
 
\begin{tcolorbox}
\emph{\textbf{Summary for RQ1:}} 
The popular packages measured by the number of monthly downloads in the two SCs cover 34 domains belonging to eight categories. 
Over 85\% of packages in either SC fall into \emph{Applications}, \emph{Infrastructure}, and \emph{Sciences} categories. 
PyTorch SC provides a higher percentage of \emph{Applications} packages such as \emph{NLP} and \emph{CV}. TensorFlow SC favors \emph{Infrastructure} packages such as \emph{MLOps Platform} and \emph{Deployment}, suggesting that TensorFlow SC and PyTorch SC have developed competitive advantages in \emph{Applications} and \emph{Infrastructure} domains, respectively. 
\end{tcolorbox}

\section{RQ2: \rqtwo} 

\subsection{Method}\label{ss: rq2 method}
Conceptually, SC is a directed graph, enabling us to employ graph community detection techniques to decompose SC's complex structure into package clusters. 
First, in this RQ, we want to evaluate the impact of the derived packages in the SC. Therefore, the root nodes (i.e., TensorFlow and PyTorch framework) and their connecting edges are removed from further analysis, leaving us with 892 nodes with 938 edges in the TensorFlow SC, and 2,044 nodes with 2,661 edges in the PyTorch SC. 
A considerable number of nodes (65.2\% in TensorFlow SC and 37.6\% in PyTorch SC) are isolated points, suggesting that their only dependency in the SC is the DL framework. 
Next, we run the Leiden community detection algorithm on the pruned SCs. This algorithm is chosen for its efficiency, stability, and capability of forming dense clusters, which captures the structural characteristics of SC well~\cite{leiden}. 

We aim to understand how packages engage in the SC and identify critical packages that attract other packages to the SC. To that end, we focus on the shape and complexity (measured by two widely used metrics~\cite{ecsaw16-alexandredecan, emse19-decan}: \emph{cluster size}, i.e., the number of nodes, and \emph{average degree}, i.e., the number of edges over the number of nodes) of the cluster. They demonstrate the distribution of the packages in the whole picture, which provides a basic understanding of the SC's characteristics. Specifically, the cluster's shape describes how packages within the cluster are interdependent overall, and the average degree of the cluster indicates the average number of dependency relationships of packages in the cluster. They can well reflect how packages engage in the SC through dependencies. The cluster size refers to the number of packages in the cluster and helps to identify packages that increase package engagement in the SC. 

To categorize clusters by shapes, we first visualize all clusters as figures\footnote{Please refer the figures/TensorFlow folder and figures/PyTorch folder in the replication package.} with \code{igraph}, a package providing common graph analysis tools~\cite{igraph}. 
Next, the same inspectors as Section~\ref{ss:rq1 method} conduct open coding~\cite{icse16-stol} to determine cluster shapes. 
Specifically, they observe how packages are interdependent in the clusters' visualization figures repeatedly. Then one inspector assigns each cluster to an initial code that describes the dependency structure among packages in the cluster and discusses with the other inspector to iteratively refine the codes until reaching an agreement. After that, they together develop a coding guide. 
To verify the reliability, they conduct independent coding on all clusters with the coding guide and the Cohen's Kappa value is 0.907. The conflicts are resolved by a meeting. The above process takes about 150 man-hours and related materials are provided in the \href{https://github.com/gaokai320/PyPI-DLSC}{replication package}.

\subsection{Results}
\begin{figure}
    \centering
    \includegraphics[width=0.7\linewidth]{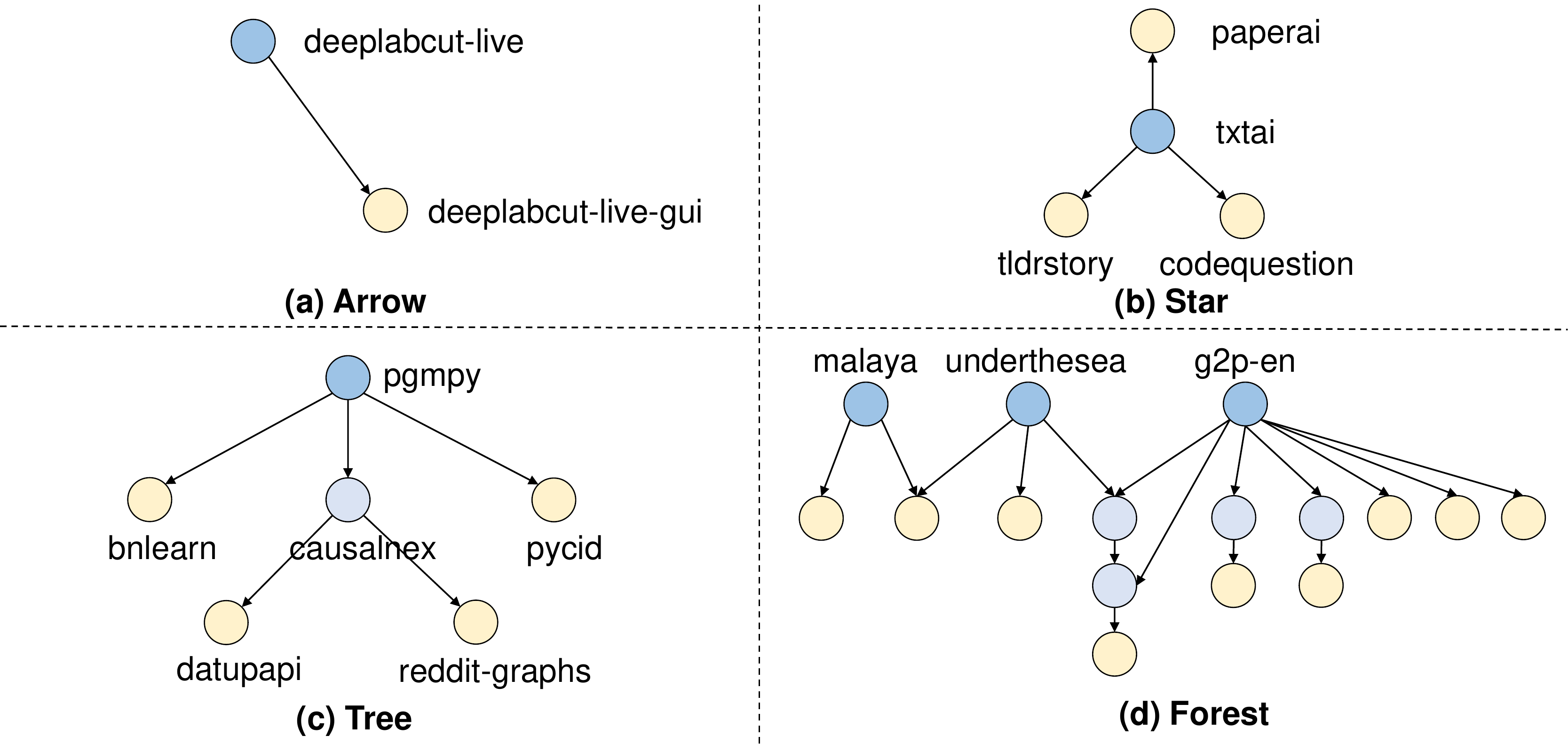}
    \caption{Example of the four cluster shapes.}
    \label{fig:shape example}
\end{figure}

\begin{figure}
    \centering
    \includegraphics[width=0.7\linewidth]{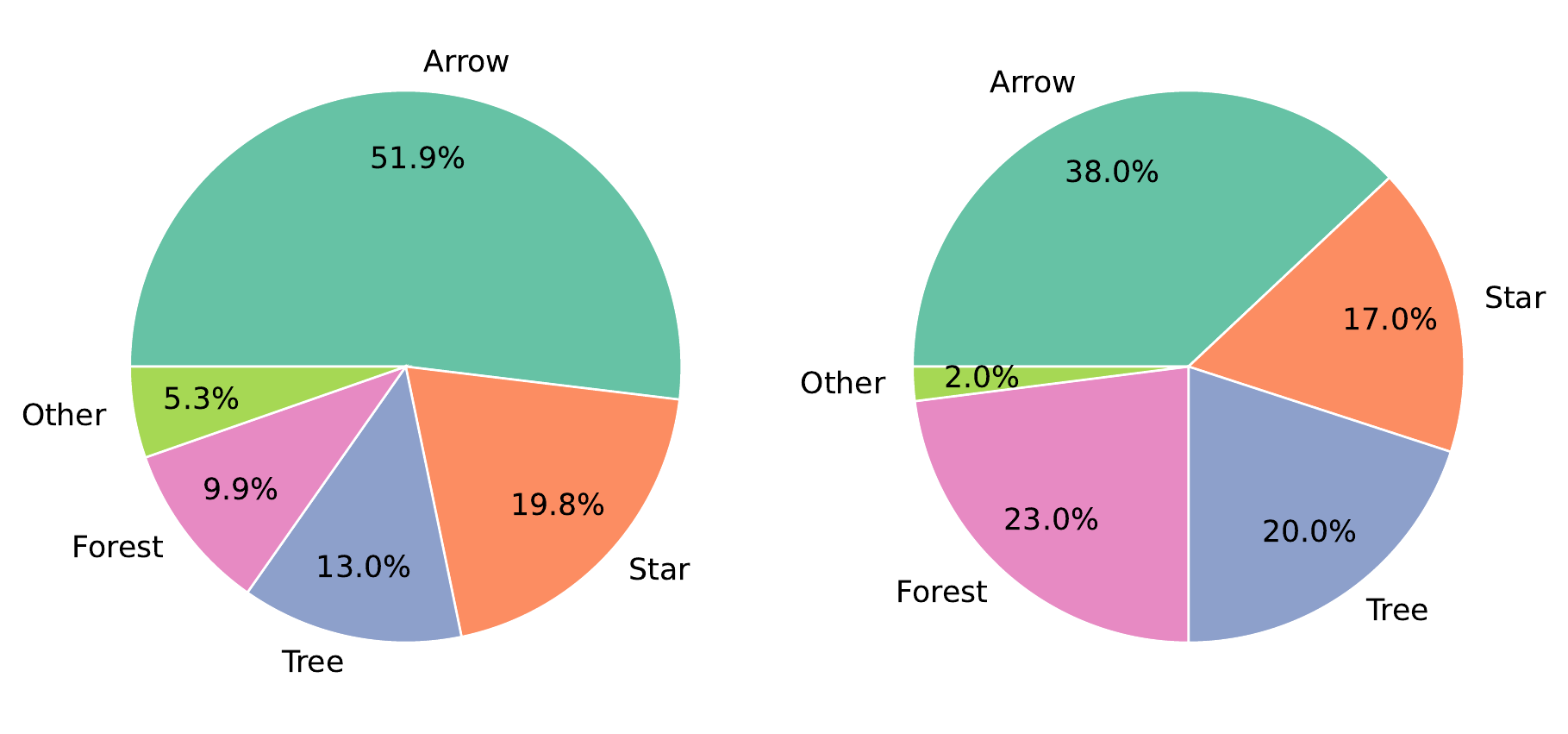}
    \caption{Distribution of cluster shapes in TensorFlow SC (left) and PyTorch SC (right).}
    \label{fig:shape dist}
\end{figure}

\textbf{Cluster Shape.} 
The Leiden algorithm detects 131 and 100 package clusters in the pruned TensorFlow and PyTorch SC respectively. 
The clusters mainly exhibit four shapes: \textbf{Arrow}, \textbf{Star}, \textbf{Tree}, and \textbf{Forest}. 
Figure~\ref{fig:shape example} illustrates the four shapes and Figure~\ref{fig:shape dist} presents the proportion of each cluster shape in the two SCs. 

\begin{itemize}[leftmargin=*,topsep=0pt]
    \item \textbf{Arrow} cluster contains only two packages with one depending on the other. 
    For example, in an Arrow cluster of TensorFlow SC, package \code{deeplabcut-live-gui} depends on package \code{deeplabcut-live} and provides GUI for it (Figure~\ref{fig:shape example}(a)). 
    68 (51.9\%) and 38 (38.0\%) clusters in the two SCs are in this shape and take 15.2\% and 3.7\% of packages respectively. Notably, packages in these clusters do not depend on packages in other clusters. 
    \item \textbf{Star} cluster presents a centralized shape that contains at least three packages with a package dependent by the remaining packages. In Figure~\ref{fig:shape example}(b), the package \code{txtai} is directly dependent by the remaining three packages. 
    26 (19.8\%) and 17 (17.0\%) clusters are in this shape and account for 12.1\% and 3.3\% of packages in the two SCs. Most (23 and 14) Star clusters contain no more than five packages. 
    \item \textbf{Tree} cluster presents a hierarchy shape with a \emph{root package} on which the other packages in the cluster depend directly or transitively. Compared with Star clusters, transitive dependencies exist among packages in Tree clusters. 
    In Figure~\ref{fig:shape example}(c), \code{pgmpy} is the cluster's root package that is directly or transitively dependent by the remaining five packages. 
    17 (13.0\%) and 20 (20.0\%) clusters are in this shape and take 26.8\% and 38.6\% of packages respectively. 
    Specifically, the largest cluster in either SC is in this shape. 
    \item \textbf{Forest} cluster shows a more complex hierarchy shape with more than one root package on which the other packages in the cluster depend. 
    Figure~\ref{fig:shape example}(d) shows a Forest cluster in TensorFlow SC. This cluster contains three root packages, i.e., \code{malaya}, \code{underthesea}, and \code{g2p-en}, dependent by the remaining 13 packages in the cluster. 
    13 (9.9\%) and 23 (23.0\%) clusters are in this shape and take 43.9\% and 54.3\% of packages respectively. Over 75\% of Forest clusters contain more than 10 packages in the two SCs. 
\end{itemize}

In addition to the above four shapes, very few (17 and 2) packages form clusters in \textbf{Other} Shape. 
Specifically, three clusters in TensorFlow SC and two clusters in PyTorch SC contain only one package whose later version depends on their early version. One cluster in TensorFlow SC contains two packages that depend on each other. Three clusters in TensorFlow SC contain a package directly depending on the rest packages. 
Since the number of Other clusters is small and they contain very few packages, in the following, we focus on the four main shapes.

\begin{figure}
    \centering
    \includegraphics[width=0.7\linewidth]{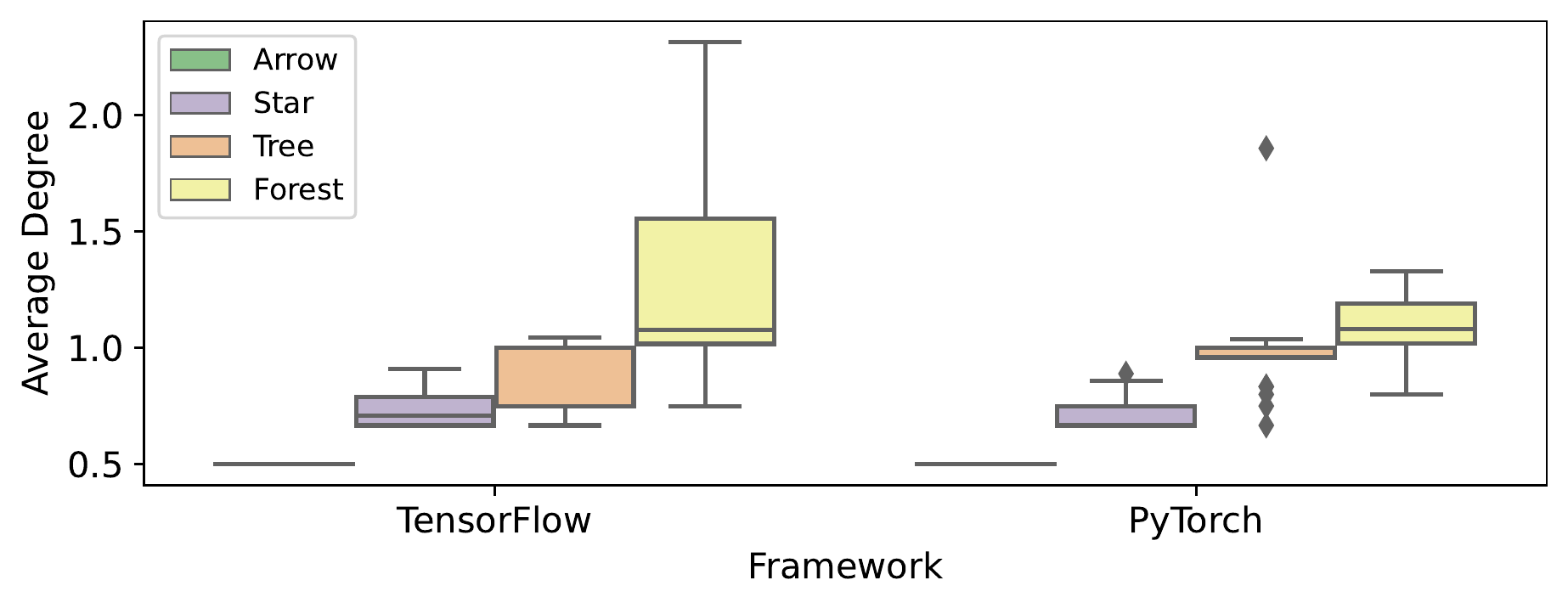}
    \caption{Distribution of the average degree.}
    \label{fig:cluster shape}
\end{figure}

\textbf{Cluster Complexity.} 
To understand the complexity of different cluster shapes, we compare their sizes and average degrees. 
Firstly we apply the Mann–Whitney U test~\cite{mwutest} to compare the sizes of Arrow, Star, Tree, and Forest clusters. Since we perform multiple tests, we use the Holm–Bonferroni method to adjust the p-values to control the family-wise error rate~\cite{holm}. The results suggest that in the two SCs Forest clusters contain more packages than Tree clusters (adjusted p-value: 0.009 and 1.40e-4) and Star clusters contain more packages than Arrow clusters (adjusted p-value: 6.57e-21 and 1.32e-12). However, the sizes of Tree clusters and Star clusters are not significantly different (adjusted p-value: 0.188 and 0.112). 
Figure~\ref{fig:cluster shape} presents the distribution of the average degree for different shapes of clusters (each data point represents the average degree of a cluster). We can observe that the average degree of Arrow, Star, Tree, and Forest clusters increases gradually with the median as 0.50, 0.71, 1.00, 1.08 for TensorFlow SC and 0.50, 0.67, 1.00, 1.08 for PyTorch SC. The Mann–Whitney U test with Holm–Bonferroni method also suggests a significant difference. 
Particularly, the average degree for most Forest clusters (11 of 13 and 20 of 23) is no less than one, indicating that on average a package in the Forest cluster has dependency relations with at least two packages. 
We also investigate the depth of Tree and Forest clusters considering their complexity. The cluster's depth is defined as the length of the longest path between root packages and other packages in the cluster and reflects the ability of root packages to attract transitive dependents. Overall, the average depth of Forest clusters is higher than Tree clusters in either SC (TensorFlow SC: 2.46 vs. 2.06, PyTorch SC: 2.39 vs. 2.30), suggesting that root packages in Forest clusters are more likely to attract transitive dependents. 
To conclude, Arrow, Star, Tree, and Forest clusters show increasingly complex dependency relations. 
The proportion of Tree and Forest clusters and the proportion of packages in Tree and Forest clusters in PyTorch SC (43.0\%, 92.9\%) is much higher than that in TensorFlow SC (22.9\%, 70.7\%). It suggests that PyTorch SC has more complex dependency relations among packages and has a more centralized structure. Two reasons may account for this. First, PyTorch SC has a higher percentage (429, 13.1\%) of packages with dependents than TensorFlow SC (276, 10.8\%). Second, packages in PyTorch SC tend to depend on more packages than in TensorFlow SC, as suggested by the Mann-Whitney U test (p-value: 2.17e-99). 

To gain more insights into SC's structural characteristics, we further inspect large clusters with more than 10 packages in each SC, resulting in 15 clusters in TensorFlow SC and 20 clusters in PyTorch SC. We find eight (of 15) and 18 (of 20) large clusters in the shape of Forest in TensorFlow SC and PyTorch SC respectively. 
We also inspect the package with the most dependents in these large clusters (core packages for short). 
One notable thing is that the core packages for 1/3 of large clusters in TensorFlow SC and 1/4 of large clusters in PyTorch SC are official packages, i.e., packages released by the DL framework team. It suggests that official packages play an important role in forming the SC. 
We also find that these core packages can well reflect the differences in the domain distribution of the two SCs revealed in RQ1. First, five (of the 20) core packages in PyTorch SC are about \emph{CV} but none of the 15 core packages in TensorFlow SC are about \emph{CV}, which coincides with the huge difference in the proportion of \emph{CV} packages in the two SCs. Second, 40\% of the 15 core packages in TensorFlow SC fall into the \emph{Infrastructure} category, and the percentage in PyTorch SC is 25\%, which is consistent with the higher percentage of \emph{Infrastructure} packages in TensorFlow SC. 

\begin{tcolorbox}
\emph{\textbf{Summary for RQ2:}} 
The Leiden algorithm detects 131 and 100 clusters in TensorFlow and PyTorch SC respectively. 
The clusters mainly present four shapes: Arrow, Star, Tree, and Forest, with increasing dependency complexity. 
\textit{Tree} and \textit{Forest} clusters account for about 22.9\% and 43.0\% of clusters but contain the majority (70.7\% and 92.9\%) of packages, indicating that a small number of packages attract most packages to the SC. The higher percentage of Tree and Forest clusters in PyTorch SC suggests that PyTorch SC has more complex dependency relationships among packages than TensorFlow SC. The core packages of large clusters tend to be official packages in either SC and demonstrate the (targeted) specialty of each SC.
\end{tcolorbox}

\section{RQ3: \rqthree}\label{ss: dynamic}

\subsection{Method}
We consider a package to have disengaged from a PyPI DL SC if its latest version in the SC ($V_{sc}$) is earlier than its latest version in PyPI ($V_{pypi}$). 
In other words, it removes the DL framework and its dependents from installation dependencies after version $V_{sc}$. 
In this way, we find that 364 and 201 packages have disengaged from TensorFlow and PyTorch SC respectively. 
We identify disengagement reasons from the package's code repository since the code repository documents rich development activity data (e.g., commits, issues, pull requests, and release notes~\cite{icpc22-wujianyu, saner23-wujianyu}) that have been used to identify reasons for various software development scenarios in prior work, e.g., library migration~\cite{fse21-hehao} and Dependabot deprecation~\cite{TSE2023-He}. 
Specifically, we take three steps as follows: 

\circled{1} \emph{Locate packages' code repository.} For each package that has disengaged from the SC, we first identify its code repository from the repository link on its PyPI page following prior work~\cite{fse21-sotovalero, fse21-vu, ase21-vu}. If the link is broken or not present, we Google its name and description for its code repository. About 75\% of the packages are linked to their repositories in either SC (TensorFlow SC: 275/364, PyTorch SC: 150/201). The ratio is comparable with previous work~\cite{fse18-maratvaliev}. 

\circled{2} \emph{Extract disengagement-relevant text.} 
For each located code repository, we first identify the commit that removes the DL framework and its dependents from the dependency specification files such as \code{setup.py} and \code{requirements.txt}. If the commit message justifies the disengagement (i.e., removal of dependencies), we record relevant texts. Otherwise, we go through the issues, pull requests, and release notes that contain the name of the dependency and pick relevant texts. 
Finally, we extract disengagement-relevant texts for about 30\% of packages in the two SCs (TensorFlow SC: 83/275, PyTorch SC: 44/150). 
The ratio is low because developers do not document the rationale behind their work, as revealed in prior work~\cite{icse22-zhangyuxia}. Despite this, the ratio is comparable with prior work that identified migration reasons~\cite{fse21-hehao} and Dependabot deprecation reasons~\cite{TSE2023-He} from code repositories. 

\circled{3} \emph{Conduct thematic analysis.} 
The first two authors conduct a thematic analysis on the extracted texts to mine reasons for package disengagement. 
Each inspector independently reads and re-reads the extracted texts, generates initial codes describing disengagement reasons, and merges similar codes into themes. 
Then they discuss the generated codes and themes until they reach an agreement and develop a coding guide. 
Next, each inspector independently uses the coding guide to code all extracted texts. 
Note that multiple codes can be assigned to one text fragment. In this case, we use MASI distance~\cite{masi} to measure the distance between the two inspectors’ codes and Krippendorff's alpha~\cite{krippendorff2011computing} to measure inter-rater reliability. Krippendorff's alpha is 0.927, which is higher than the recommended threshold of 0.8~\cite{krippendorff2018content}. 
A meeting is held to resolve conflicts and to assign the final codes. The process takes about 200 man-hours and related materials can be found in the \href{https://github.com/gaokai320/PyPI-DLSC}{replication package}.

\subsection{Results}\label{ss: rq3 results}
\begin{figure}
    \centering
    \includegraphics[width=0.8\linewidth]{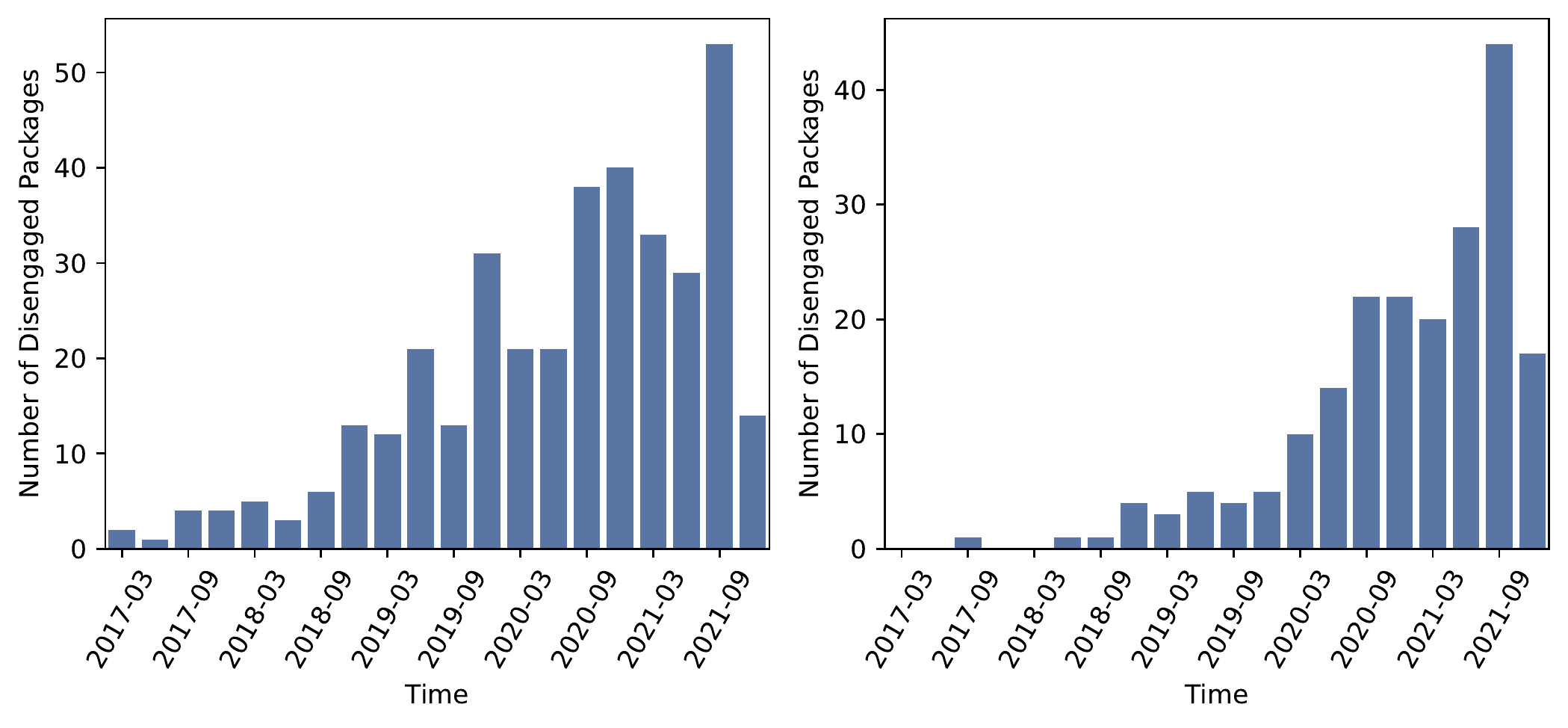}
    \caption{Number of disengaged packages each quarter in TensorFlow SC (left) and PyTorch SC (right).} 
    \label{fig:deprecation trend}
\end{figure}

Figure~\ref{fig:deprecation trend} shows the trend of disengaged packages in the two SCs. We can see that the number of packages that disengaged from either SC each quarter shows an increasing trend overall (note that we only have partial data for the last quarter of 2021). In the 3rd quarter of 2021, both TensorFlow and PyTorch SC experienced a sharp increase in disengaged packages. 
These findings indicate that TensorFlow and PyTorch SC both suffer from increasingly severe package disengagement issues. 

We identify seven disengagement reasons related to three aspects from the 83 and 44 disengaged packages in TensorFlow and PyTorch SC respectively. 
Table~\ref{tab:detach reason} shows the number and proportion of disengaged packages mentioning each reason in either SC. Note that the sum of the proportions exceeds 100\% because a package may disengage from the SC for multiple reasons.
In the remainder of this section, we will elaborate on each reason. 

\noindent\textbf{Dependency.} 38 (45.78\%) and 13 (29.55\%) package disengagements in TensorFlow and PyTorch SC are due to dependency issues. This category consists of two reasons: the removed dependency causes incompatibility or is bloated. 

\myfancylabel\;\emph{Incompatibility.} Some packages disengage from the SC because their dependencies induce incompatibility \emph{issues or errors} with other required packages, hardware, or operating systems. 
It is the most common reason in TensorFlow SC mentioned by 23 (27.71\%) disengaged packages. The predominant cause is the conflicts among TensorFlow packages. 
As mentioned in Section~\ref{s: sc construction}, the TensorFlow framework releases three packages: \code{tensorflow}, \code{tensorflow-cpu}, and \code{tensorflow-gpu}. 
The three packages share the same API names, meaning that installing one of them overrides the other already installed package.\footnote{\url{https://github.com/tensorflow/tensorflow/issues/7166}} 
Developers often adopt two strategies to resolve this issue. 
The first is to remove TensorFlow packages from dependencies and guide end users to install the correct package for their environments. 
The other is to check the environment (e.g., the presence of GPUs) in \code{setup.py} and install the corresponding TensorFlow package. 
The downside of this strategy is that the package must be installed from the source.

\begin{table}[t]
  \caption{Reasons for Package Disengagement.}
  \label{tab:detach reason}
  \renewcommand{\arraystretch}{1.2}
  \begin{threeparttable}
  \begin{tabular}{lrr}
  \toprule
    Reason & TensorFlow SC\tnote{1} & PyTorch SC\tnote{2}\\
    \midrule
    \textbf{Dependency} & 38 (45.78\%) & 13 (29.55\%) \\
    \MyIndent Incompatibility & 23 (27.71\%) & 10 (22.73\%) \\
    \MyIndent Bloated & 15 (18.07\%) & 4 (9.09\%) \\
    \textbf{Functionality} & 27 (32.53\%) & 24 (54.55\%) \\
    \MyIndent Performance & 16 (19.28\%) & 7 (15.91\%) \\
    \MyIndent Simplification & 9 (10.84\%) & 13 (29.55\%) \\
    \MyIndent Framework-Independent & 3 (3.61\%) & 5 (11.36\%) \\
    \textbf{Installation} & 25 (30.12\%) & 14 (31.82\%) \\
    \MyIndent Flexibility & 15 (18.07\%) & 1 (2.27\%) \\
    \MyIndent Size Trimming & 10 (12.05\%)  & 13 (29.55\%) \\
  \bottomrule
  \end{tabular}
  \begin{tablenotes}
    \footnotesize
    \item[1] 83 disengaged packages in total in TensorFlow SC.
    \item[2] 44 disengaged packages in total in PyTorch SC.
    \end{tablenotes}
  \end{threeparttable}
\end{table}

\myfancylabel\;\emph{Bloated.} 15 (18.07\%) and four (9.09\%) packages disengage from the two SCs because they do not use the removed dependencies at all, or package developers assume that the dependencies have been already installed in the environment. For example, a package that has removed PyTorch from its dependencies explained that \emph{If you set up the development environment to work on PyTorch, you would probably have it installed already.}\footnote{\url{https://github.com/Quansight/pytest-pytorch/issues/13\#issue-861367114}} 
Bloated dependencies increase the package's installation size and import time. Once found, developers will remove them from dependency specification files. 
It appears that the bloated dependency issue is more common among packages in TensorFlow SC, which is consistent with the manual inspections in Section~\ref{s: sc construction}. 

\noindent\textbf{Functionality.} 
27 (32.53\%) and 24 (54.55\%) packages disengage from TensorFlow and PyTorch SC respectively for functionality improvements. 

\myfancylabel\;\emph{Performance.} Sixteen and seven packages disengage from TensorFlow and PyTorch SC for better performance, e.g., simpler model, faster import time and execution time. 
Some packages just remove dependencies, for example, \emph{Replace MTCNN with Mediapipe to remove TensorFlow + Keras dependency}.\footnote{\url{https://github.com/apangasa/bestof/issues/21}} 
Some choose to migrate to other packages, e.g., \emph{The code may become cleaner and faster and more flexible if we strip out PyTorch, and instead (maybe) use concurrent.futures.ProcessPoolExecutor.}\footnote{\url{https://github.com/openclimatefix/nowcasting_dataset/issues/86}} 

\myfancylabel\;\emph{Simplification.} 
Some packages provide cumbersome functionalities depending on a variety of packages. 
They disengage from the SC to simplify their functionalities for developers easier to use. 
Nine (10.84\%) and 13 (29.55\%) disengaged packages in TensorFlow and PyTorch SC mention this reason. This reason together with \emph{Size Trimming} is the two most common disengagement reasons in PyTorch SC. 
Packages usually take three approaches to simplify their functionalities. 
Some packages turn certain functionalities into optional features and the dependencies required by these functionalities into extra dependencies~\cite{PEP508}. For example, \emph{These ... may not be needed if the user needs to explain sklearn models. If needed, they can installed as optional dependencies using pip, pip install dice-ml[deeplearning]}.\footnote{\url{https://github.com/interpretml/DiCE/pull/74}} 
Some packages split their functionalities into multiple packages. For example, \emph{We have recently migrated the pipeline deeplearning-prepare-data to our sister project ClinicaDL which handles everything related to deep-learning.}\footnote{\url{https://github.com/aramis-lab/clinica/issues/458}} 
Some packages take a rather aggressive approach of completely removing functionalities that require these dependencies. For example, \emph{we will remove all dependencies and code related model training and experiments to keep datasetinsights focused on io and statistics.}\footnote{\url{https://github.com/Unity-Technologies/datasetinsights/issues/143}} 

\myfancylabel\;\emph{Framework-Independent.} 
Some packages disengage from a particular DL SC to support more DL frameworks. 
For example, \emph{Tonic is now free from direct PyTorch and PyTorch Vision dependencies, meaning that if someone wanted to use it with another pipeline (e.g. TensorFlow), they would be able to do so.}\footnote{\url{https://tonic.readthedocs.io/en/latest/about/release_notes.html\#id4}} 
Three and five disengaged packages in the two SCs mention this reason. 

\noindent\textbf{Installation.} 
25 (30.12\%) and 14 (31.82\%) packages disengage from TensorFlow and PyTorch SC for two reasons related to installation. 

\myfancylabel\;\emph{Flexibility.} Packages disengage from the SC to ease and simplify the installation procedure to adapt to various environments. 
The difference between \emph{Flexibility} and \emph{Compatibility} is that the latter is passively triggered by errors while the former is proactively performed to relax installation requirements. For example, \emph{There are multiple official and community TensorFlow packages out there ... The main tensorflow package works in most cases but ultimately we should allow users to install an alternative package.}\footnote{\url{https://github.com/OpenNMT/OpenNMT-tf/pull/802}} 15 disengaged packages in TensorFlow SC and only one disengaged package in PyTorch SC mention this reason. 
As discussed above, TensorFlow provides separate packages for different hardware platforms, whereas PyTorch provides only a single package. Therefore, packages in TensorFlow SC care more about hardware requirements. 

\myfancylabel\;\emph{Size Trimming.} 
Either TensorFlow or PyTorch is large whose built distributions occupy hundreds of megabytes of space, resulting in a large installation size for packages in the SC as well. Therefore, some packages disengage from the SC to reduce their installation size. 
Different from \emph{Bloated} where packages disengage from the SC because they do not use the dependencies, dependencies are used by disengaged packages in the case of \emph{Size Trimming}. 
In addition to removing or turning dependencies into extra dependencies, we also find packages that choose to migrate to ONNX in either SC. For example, 1) \emph{The pytorch dependency is huge ... Let's try to replace it by ONNX, which is hopefully smaller.}\footnote{\url{https://github.com/raymon-ai/raymon/issues/50}} 2) \emph{switched from tensorflow to onnx runtime, smaller binary size.}\footnote{\url{https://github.com/paradigmn/ultrastar_pitch/issues/5}}. 
10 (12.05\%) and 13 (29.55\%) packages disengage from the two SCs to trim their installation size, suggesting packages in PyTorch SC are more likely to suffer from large installation sizes. 

We further analyze the trends of the seven reasons to discern why the number of disengaged packages has increased in the two SCs (as shown in Figure~\ref{fig:deprecation trend}). We find that simplifying functionality and trimming installation size are the major reasons for the increase of disengaged packages in PyTorch SC, possibly due to the increasing size of the PyTorch framework and the increasingly cumbersome functionalities provided by these disengaged packages. We also find that increasing packages disengage from TensorFlow SC due to incompatibility and performance issues, possibly because more and more packages encounter conflicts among the three TensorFlow packages and TensorFlow fails to meet the disengaged packages' performance requirements such as import time and state-of-the-art models. 

\begin{tcolorbox}
\emph{\textbf{Summary for RQ3:}} 
The number of disengaged packages in either SC shows an increasing trend. 
Packages disengage from the SC for seven reasons related to three aspects: dependency issues, functional improvements, and ease of installation. 
The most common disengagement reason in TensorFlow SC is dependency incompatibility issue, and that in PyTorch SC is to simplify functionalities and reduce installation size, indicating packages in different DL SCs face different dependency management problems. 
\end{tcolorbox}

\section{Implications}
Our study provides rich suggestions for DL framework vendors, researchers, and stakeholders on the maintenance and dependency management practices of PyPI DL SCs. 
To elicit the practitioners' opinions on our suggestions, We interviewed five people, including two contributors to the PaddlePaddle DL framework, one contributor to the MindSpore DL framework, and two DL researchers from two large IT companies. Overall, they recognize that our suggestions are clear and informative and some suggestions coincide with their practices. They also provide additional information on some suggestions. We elaborate on these suggestions in this section.

\subsection{DL SC Maintenance}

\faTools\;\textbf{Framework improvement.} 
A decent DL framework is the foundation of a prosperous DL SC. 
Our findings reveal two issues that DL frameworks should consider. 
\begin{itemize}[leftmargin=*,topsep=0pt]
    \item \emph{Package size}. DL frameworks are very large packages whose pre-built distributions occupy hundreds of megabytes of space. As a result, installing DL packages that directly or transitively depend on them consumes a considerable amount of disk space and network traffic. 
    Many packages disengage from DL SCs, especially PyTorch SC, to reduce their installation size (RQ3). Therefore, DL frameworks may consider controlling their package size to retain packages in their SCs. We envision that DL frameworks may split into a base package with many plugin components. The base package contains commonly used functionalities such as CNN layers while the plugin components contain functionalities for specific tasks, e.g., visualization. The PaddlePaddle contributors point out that the inclusion of multiple versions of CUDA binaries is the primary reason for the overwhelming size of DL framework packages. They further reveal that they control PaddlePaddle's package size by setting a range for supported CUDA versions. 
    \item \emph{Multi-hardware support}. A prominent disengagement reason in TensorFlow SC is the conflicts among different TensorFlow packages released for CPU and GPU respectively (RQ3). 
    Currently, there are two approaches to support multiple hardware platforms. 
    The first is to build separate packages for different hardware platforms. Most DL frameworks, e.g. Tensorflow, MxNet, and PaddlePaddle, adopt this approach. 
    The second is to provide a single package that supports a wide range of hardware platforms, as PyTorch does. 
    Both approaches have their limitations: the former leads to conflicts between different packages, while the latter increases package size. The PaddlePaddle and MindSpore contributors justify that they choose the first approach because PyPI sets a size limit (60MB) for package distributions and they have to request an increase to the distribution size limits from the PyPI maintainers.\footnote{\url{https://www.dampfkraft.com/code/distributing-large-files-with-pypi.html}} 
    The dilemma is largely due to PyPI's lack of support for uploading pre-built distributions for different hardware platforms.\footnote{\url{https://discuss.python.org/t/what-to-do-about-gpus-and-the-built-distributions-that-support-them/7125}} 
    PyTorch wheel repository utilizes local version identifiers~\cite{PEP440} to differentiate distributions for different hardware platforms.\footnote{\url{https://download.pytorch.org/whl/torch_stable.html}} However, local version identifiers are not permitted on PyPI, and even worse, the \code{packaging} package has removed the support of local version identifiers recently~\cite{packaging22.0}. 
    In summary, current Python packaging specifications pose tons of challenges to DL package maintainers from supporting multiple hardware platforms elegantly. 
    Thus, the Python community may propose adjustments to the packaging specifications (e.g. reintroducing local version identifiers or having \code{pip} understand the conflicts among packages).
\end{itemize}

\faCubes\;\textbf{Package Diversity.} DL is widely adopted in our society, motivating diverse applications and tools, as evidenced by the 34 package domains identified in RQ1. The diversity of packages in turn boosts the SC's prosperity. We recommend DL framework vendors diversify the chain from two aspects. 
\begin{itemize}[leftmargin=*,topsep=0pt]
    \item \emph{Competitive advantages}. It's noteworthy that both TensorFlow and PyTorch SC have developed their specializations on \emph{Infrastructure} and \emph{Applications} packages respectively (RQ1). We also find that official packages play an important role in forming the SC (RQ2). 
    Therefore, for emerging DL frameworks, they can start by providing official packages for domains that account for the most packages in the two SCs such as \emph{NLP} and \emph{CV}. These official packages could provide common datasets and model architectures. However, we notice that not all DL frameworks prioritize CV and NLP-related packages. For example, MindSpore does not provide these packages (i.e., \code{MindCV} and \code{MindNLP}) until 2023. 
    Then they may spend efforts on specific domains (e.g., Social Aspects and Reinforcement Learning) to increase their popularity and unique competitiveness in the DL community. This suggestion coincides with what MindSpore does as confirmed by the MindSpore contributor. Specifically, MindSpore has provided many \emph{Sciences} packages such as \code{MindEarth} and \code{MindSPONGE} to gain a competitive advantage in the \emph{Sciences} domains. 
    \item \emph{Emerging areas}. DL framework vendors may track the emerging and promising areas in which the adoption of DL techniques makes a difference. 
    For example, as SE researchers, we are surprised at the absence of SE-related popular packages in either SC. Given that DL techniques are increasingly applied in diverse SE tasks (e.g., code search~\cite{ase18-chenqingying} and code generation~\cite{aaai20-sun}) and show promising results, we anticipate in the near future that both DL and SE communities could together build tools to facilitate the development of DL for SE, e.g., the ``ImageNet'' of SE, common data (e.g., abstract syntax tree) preprocessing tools, and large pre-trained models for various SE tasks. The PaddlePaddle and MindSpore contributors recognize the deficiency of AI4SE packages and consider devoting more efforts to this area. 
\end{itemize}

\faHandshake\;\textbf{Collaboration.} 
First, DL is widely adopted in various scientific disciplines as revealed by RQ1. Applying DL techniques to these areas usually requires domain expert knowledge. Therefore, DL framework vendors can establish collaboration with researchers and practitioners from other disciplines. 
Second, DL models are growing dramatically in their size, involving billions of parameters, hundreds of gigabytes of training data, and large GPU clusters. 
Training such models is barely affordable for any single organization and necessitates collaboration with other organizations. 
Finally, we find some packages depend on packages from both SCs (RQ1), which reflects the lack of tool support in one particular SC. 
Therefore, collaborations with packages in other DL SCs to complement tool support are also strongly encouraged. The PaddlePaddle contributors also emphasize the collaboration between DL framework vendors and hardware vendors to facilitate the deep fusion of hardware and software and to fully unlock the computing potential of the hardware. 

\faExclamationTriangle\;\textbf{Risk management.} 
Tree and Forest clusters assemble about 70\% $\sim$ 90\% packages through their root packages (RQ2), indicating dense dependency relationships among most DL packages. 
Such relationships magnify SC risks, e.g. malicious code injections and vulnerabilities in a popular root package trigger the panic of maintainers of a large number of its dependents. Therefore, for DL SCs and SCs in other fields and packaging ecosystems, more attention should be paid to these root packages to avoid and reduce risks. 
On the one hand, vulnerabilities in root packages will propagate along the SC and impact other packages within the same cluster. 
Therefore, vulnerability notification and patch propagation tools can be designed to mitigate the cascading impact of security issues of root packages. Such tools may integrate with package managers (e.g. \code{npm audit}\footnote{\url{https://docs.npmjs.com/cli/v9/commands/npm-audit}}) or code hosting platforms (e.g. Dependabot\footnote{\url{https://github.com/dependabot}}) to offer timely update notifications to downstream developers and end users. 
On the other hand, if root packages disengage from the SC, i.e., remove dependencies on the DL framework and its dependents, their dependents will also disengage from the SC. 
Therefore, the retention of root packages is also critical to SC's stability and sustainability. 

\subsection{DL SC Dependency Management}

\faExchange*\;\textbf{Interoperability.} 
We find TensorFlow and PyTorch SC differ greatly in the distribution of package domains (RQ1), indicating that the resources of the DL community are scattered among different DL SCs. This may add extra burden to DL developers. 
For example, if a developer uses package A in PyTorch SC for prototype development and package B in TensorFlow SC for deployment, she/he must be familiar with both frameworks and perhaps write redundant code. 
Therefore, we consider interoperability among frameworks can help developers harness the strengths of different DL SCs~\cite{fse20-liuyu}. 
Current interoperability tools such as ONNX and MMdnn~\cite{fse20-liuyu} mainly focus on converting models between different DL frameworks by defining a common set of operators. However, due to the discrepancies among operators supported by different DL frameworks, the engineering efforts are formidable and the application scenarios are still limited. The PaddlePaddle contributors add that the different and evolving implementations of the same operator by different frameworks also pose challenges to existing interoperability tools. 
Ideally, DL frameworks could conform to a common interoperability tool. However, it is not easy for DL framework vendors to reach a consensus due to different incentives~\cite{tse21-zhangyuxia}. Therefore, DL practitioners could collectively complement the support of DL frameworks for existing tools. 
Moreover, novel techniques to achieve interoperability among DL frameworks can be explored. 

\faPython\;\textbf{Dependency division.} 
DL packages usually introduce many dependencies, which increases the risk of dependency issues like dependency incompatibility, cumbersome functionalities, and large installation sizes (RQ3). 
To simplify their dependency tree and resolve the above issues, some packages split non-core functionalities that require dependencies into ``extras''. 
Specifically, these packages choose to split their functionalities into a base package with main dependencies and multiple optional components with corresponding extra dependencies they require. 
By default, only dependencies required by the base package are installed. 
Extra dependencies are installed only if the user explicitly declares the use of optional components. 
To divide dependencies, a good modular architecture is a prerequisite. That is, extra dependencies are not imported by Python code files corresponding to the functionalities provided in the base package so that the base package works properly without extra dependencies. Therefore, practices on how to properly specify extra dependencies and related automation tools can be explored. The two DL practitioners consider this suggestion very useful since they have developed packages supporting different frameworks and this suggestion can help them better organize dependencies and free users from downloading unnecessary dependencies. We consider that similar practices can be applied to SCs in other packaging ecosystems. 

\faFilter\;\textbf{Dependency debloating.} 
We find 15 and four packages that have disengaged from TensorFlow SC and PyTorch SC mentioned that they did not use the dependencies declared in the metadata at all (RQ3), suggesting that bloated dependencies may be common among PyPI packages. 
Bloated dependencies increase the package's installation size and slow the import process. 
Although some work has investigated bloated dependencies in other ecosystems like Java/Maven, e.g., debloating tools~\cite{fse20-bruce} and evolution~\cite{emse21-sotovalero, fse21-sotovalero}, bloated dependencies in Python/PyPI and other packaging ecosystems is still under-investigated~\cite{tse22-caoyulu} and tools for automatically debloating Python dependencies is lacking. 
We recommend researchers bridge these knowledge gaps in the future. 

In addition to DL frameworks, we consider that the implications of controlling package size, risk management, dependency division, and dependency debloating are also applicable to packages in other packaging ecosystems such as NPM. Take the implication of controlling package size as an example, NPM developers often suffer from failures in publishing large packages to NPM and it is recommended that large packages \emph{``should be avoided or broken down into smaller packages''}.\footnote{\url{https://github.com/npm/npm/issues/12750}}

\section{Threats to Validity} 
\textbf{Internal validity} concerns how we draw our findings. The main threat comes from author bias introduced by manual labeling of package domains, cluster shapes, and disengagement reasons. 
To mitigate author bias, two inspectors perform labeling independently, develop coding guides through a series of discussions, and measure the inter-rater reliability throughout the process. The inter-rater reliability results all indicate a high agreement. 
Then, extracting dependency relationships between packages from the package's distribution metadata is a standard practice in many recent studies (e.g., ~\cite{icse22-liuchengwei, emse19-decan}). 
However, not all runtime dependencies are listed in the distribution metadata. 
To evaluate how severe this case is, we randomly sample 100 packages in the two SCs, download their latest distributions, and compare packages imported in the code with dependencies declared in the metadata. We find that six packages import packages that are not declared in the metadata, but all six packages do not have dependent packages and have a monthly download of less than 900. 
Thus, we believe the impact of this case should be minor and will not invalidate our results. 
Future research may analyze the packages' source code for a more precise dependency resolution. 
The selection of popular packages for domain distribution analysis also poses threats to internal validity. To evaluate the representativeness of the selected packages, we further sample 334 and 344 packages in TensorFlow and PyTorch SC respectively (95\% confidence level and 5\% confidence interval). We use the derived domains to label the sampled packages and obtain similar findings: over 85\% (TensorFlow SC: 89.6\%, PyTorch SC: 87.4\%) of packages in either SC fall into \emph{Applications}, \emph{Infrastructure}, and \emph{Sciences} categories; the two SCs favor \emph{Infrastructure} (TensorFlow SC: 32.9\%, PyTorch SC: 22.2\%) and \emph{Applications} (TensorFlow SC: 39.4\%, PyTorch SC: 51.2\%), respectively. Therefore, we believe that the selected popular packages are representative. 
Only about 30\% of disengaged packages are used to identify disengagement reasons, which poses another threat. Although we do not know how representative these packages are, packages whose code repository documents the rationale behind disengagement may be of better quality than those without. So their practices should also provide valuable insights into package disengagement. 
Finally, the use of not-so-recent data may threaten our findings since our findings may not reflect the recent progress of DL such as large language models (LLMs) and diffusion models. If thinking it carefully, we can see the use of not-so-recent data mainly threatens the findings of RQ1. However, we argue that their impact on our findings (especially RQ1) is limited for three reasons: 1) The progress (e.g., LLMs) is mostly shared via pre-trained model weights in the Hugging Face platform instead of software packages. 2) The underlying techniques behind them have already been proposed before 2021 and are provided by packages in our dataset. For example, the transformer architecture is widely used by LLMs and was proposed in 2017. Our dataset contains many packages providing transformer architectures such as \code{pytorch-transformers}. 3) The progress mainly concentrates on the NLP and CV domains, which are the two most popular domains in our dataset. It suggests that the domain distribution may not change greatly even taking the progress into account. Therefore, despite the use of not-so-recent data, we believe our findings still provide valuable insights into DL package SCs. 

\textbf{External validity} concerns the generalization of our findings. 
This paper investigates the SCs of the two most popular and representative DL frameworks, TensorFlow and PyTorch. 
As a comparison, we also construct SCs for another three DL frameworks, i.e., MindSpore, PaddlePaddle, and MxNet, which include 6, 33, and 87 packages respectively, far less than the number of packages in TensorFlow SC (2,567) and PyTorch SC (3,278). Therefore, we believe the results obtained from the two prosperous DL SCs should also provide valuable insights for other DL frameworks. 
Still, considering the unique characteristics of the DL field (e.g. the diverse hardware platforms) and PyPI packaging ecosystem, researchers must be cautious before applying our findings to other fields (e.g. web frameworks) or other packaging ecosystems (e.g. Maven or NPM). 
However, our SC construction approach and analysis framework should be generalizable to other fields and other packaging ecosystems. 
Further investigations can be conducted to understand the similarities and specifics among SCs in different fields and different ecosystems. 
Then, PyPI does not cover all Python packages since not every code repository would release packages to PyPI. However, considering that PyPI is the largest and de facto package registry for developers to find, install, and publish Python packages and hosts over 488 thousand packages to date, we believe that package SCs in PyPI are representative. Taking packages released in other platforms (e.g., GitHub code repositories) into account will lead to a more complete package SC, but it needs nontrivial efforts and is beyond the scope of the current study, e.g., how to collect as complete Python code repositories as possible, how to accurately identify packages released by a code repository, and how to identify dependents of these packages. We leave it as future work.

\textbf{Construct validity} is the degree to which our metrics of the cluster size and average degree measure the complexity of clusters. The two metrics are widely used in prior work on analyzing SCs in packaging ecosystems, e.g., (~\cite{ecsaw16-alexandredecan, emse19-decan}). We believe that the two metrics have a high potential to reflect the cluster's complexity. 

\section{Conclusion}
Through analyzing nearly six million PyPI package distributions, we construct version-sensitive SCs of TensorFlow and PyTorch and analyze their package domains, clusters, and disengagement. 
We find the popular packages in the two SCs cover 34 domains spanning eight categories and the two SCs have developed their specializations on \emph{Infrastructure} and \emph{Applications} respectively. 
We find that package clusters in the two SCs are mainly in the shape of Arrow, Star, Tree, and Forest, with increasing dependency complexity. Most packages are concentrated in Tree and Forest clusters. 
We also find packages disengage from the SC for seven reasons in three groups: dependency issues, functionality improvements, and ease of installation. The most common disengagement reasons in TensorFlow SC and PyTorch SC are different. 
Overall, the two SCs share similar package domains and cluster shapes but differ in their specialized packages and dependency management issues. 
Our findings provide rich implications for DL framework vendors, researchers, and practitioners on the maintenance and dependency management practices of PyPI DL SCs.

\begin{acks}
This work is sponsored by the National Natural Science Foundation of China 61825201 and 62332001.
\end{acks}

\bibliographystyle{ACM-Reference-Format}
\bibliography{references}

\appendix

\end{document}